\documentclass[preprint,aps,amssymb,showpacs,superscriptaddress,
nofootinbib]{revtex4}

\usepackage{graphicx}

\newcommand{\be}{\begin{eqnarray}}
\newcommand{\ee}{\end{eqnarray}}
\newcommand{\bes}{\begin{eqnarray*}}
\newcommand{\ees}{\end{eqnarray*}}

\newcommand{\ph}{\phi}

\newcommand{\la}{\lambda}

\begin{document}

\begin{flushright}
IMSc/2004/08/31 \\
gr-qc/0408030
\end{flushright}


\title{Cosmology and Static Spherically Symmetric solutions in \\
$D-$dimensional Scalar Tensor Theories: Some Novel Features}

\author{Arjun Bagchi}
\email{arjun@imsc.res.in, arjun@mri.ernet.in}
\affiliation{The Institute of Mathematical Sciences \\
CIT Campus, Chennai-600 113, INDIA.}
\affiliation{(From August 2004 onwards) \\
Harish-Chandra Research Institute \\
Chhatnag Road, Jhunsi. Allahabad-211 019, INDIA.} 

\author{S. Kalyana Rama}
\email{krama@imsc.res.in}
\affiliation{The Institute of Mathematical Sciences\\
CIT Campus, Chennai-600 113, INDIA.}

\begin{abstract}

We consider scalar tensor theories in $D-$dimensional spacetime,
$D \ge 4$. They consist of metric and a non minimally coupled
scalar field, with its non minimal coupling characterised by a
function. The probes couple minimally to the metric only. We
obtain vacuum solutions - both cosmological and static
spherically symmetric ones - and study their properties. We find
that, as seen by the probes, there is no singularity in the
cosmological solutions for a class of functions which obey
certain constraints. It turns out that for the same class of
functions, there are static spherically symmetric solutions
which exhibit novel properties: {\em e.g.} near the ``horizon'',
the gravitational force as seen by the probe becomes repulsive.

\end{abstract}

\maketitle

\section{Introduction}

Scalar tensor theories of gravity, also referred to as
generalised Brans-Dicke theories, are a natural generalisation
of Einstein's general theory of relativity \cite{bd,rest,will}.
Such theories appear naturally in various contexts, for example
in Kaluza-Klein theories, low energy effective actions of string
theory, and five dimensional brane world theories
\cite{will,fm,polchinski,rs,brane}.

In these theories the gravity sector consists of the graviton
and a non minimally coupled scalar field $\phi$, the non minimal
coupling being characterised by a function of $\phi$. There may
also be a potential for $\phi$. The matter sector consists of
probes and/or matter fields of various types which, in general,
couple to the metric and to $\phi$. For the sake of simplicity,
in the following, we will not consider matter fields and,
moreover, assume that probes couple minimally to the metric
only.

In the following, we consider scalar tensor theories in
$D-$dimensional spacetime, $D \ge 4$, consisting of the metric
$g_{\mu \nu}$, a non minimally coupled scalar field $\phi$ with
its non minimal coupling characterised by a function
$\psi(\phi)$, and probes coupled minimally to the metric $g_{\mu
\nu}$ only. The probes will follow the geodesics of $g_{\mu
\nu}$ which we, therefore, refer to as the physical metric.

In this paper we obtain vacuum solutions of the scalar tensor
theories in $D-$dimensional spacetime - both cosmological and
static spherically symmetric ones - and study their properties.
These solutions generalise those in \cite{krama} for the $D = 4$
case, and are obtained as follows.  We first transform the
gravity sector to ``Einstein frame'' where the action for the
Einstein metric $g_{* \mu \nu}$ is the standard Einstein-Hilbert
action and the canonically normalised scalar field $\phi$ is
minimally coupled to $g_{* \mu \nu}$. The corresponding
equations of motion can be solved rather easily. We then
transform the solutions back to ``physical frame'' to obtain the
physical metric $g_{\mu \nu}$ and study their properties.

Note that: (i) In Einstein frame, the probes generically couple
to the scalar field also and thus experience a corresponding
force besides the gravitational force due to $g_{* \mu
\nu}$. Therefore, the probes do not follow the geodesics of
$g_{* \mu \nu}$. Clearly, if the scalar field forces are also
taken into account then the resulting motion of the probes in
the Einstein frame will be same as that in the physical frame -
where the probes couple minimally to the physical metric $g_{\mu
\nu}$, and follow its geodesics \cite{will,fm,flanagan}. (ii) If
a potential for $\phi$ and/or matter fields, even if coupled
minimally to $g_{\mu \nu}$, are present then the resulting
equations of motions are in general difficult to solve either in
the Einstein frame or in the physical frame. See, however,
\cite{v,damour,kalyan}. Hence, in this paper, we consider only
vacuum case where the potential for $\phi$ and the matter fields
are absent.

We find that, as measured by the probes which couple minimally
to $g_{\mu \nu}$, there is no singularity in the cosmological
solutions for a class of functions $\psi(\phi)$ which obey
certain constraints. It turns out that for the same class of
functions $\psi(\phi)$, there are static spherically symmetric
solutions which exhibit interesting properties: {\em e.g.} near
the ``horizon'', the gravitational force as seen by the probe
becomes repulsive. These features are likely to have novel
implications for realistic cases, as discussed at the end of the
paper.

The plan of the paper is as follows. In section II we give a
brief outline of scalar tensor theories. In section III we
obtain cosmological solutions and study their properties. In
section IV we obtain static spherically symmetric solutions and
study their properties. In section V we summarise our results
and conclude with a discussion of their relevence to realistic
cases, thereby also pointing out various issues that need to be
studied further.

\section{Scalar tensor theory:  General considerations}

In this paper we consider the following action 
\be 
S_{total} = S + S_{probe}(g_{\mu \nu}) \; 
\ee
in $D-$dimensional spacetime with $D \ge 4$. The scalar-tensor
part $S$ of the total action is given by
\be
S = - {1\over{16\pi}} \int d^Dx {\sqrt{-g}} \; 
{e^{({D-2\over2}){\psi}}} \; \left( R 
- \frac{A}{2} (\nabla{\ph})^2 + {e^{\psi}} V(\ph) \right)
\label{P action}
\ee
where ${\psi}$ is a function of scalar field $\phi$, $A = 1 -
{(D-1)(D-2)\over2} {\psi}_{\ph}^2$, and ${\psi}_{\ph} = {d\psi
\over {d\ph}}$.  The probe action $S_{probe}$ is, for example,
that of a point particle of mass $m_0$ which couples to the
metric $g_{\mu \nu}$ only and is given by 
\be
S_{probe} = -m_0 \int d \tau \sqrt{g_{\mu\nu} 
{dx^{\mu}\over{d\tau}} {dx^{\nu}\over{d\tau}}} \; . 
\ee
The spacetime would be governed by the equations in the scalar
tensor sector and the probe will test this background without
distorting it. The probe follows the geodesics of $g_{\mu\nu}$.
So, $g_{\mu \nu}$ is the physical metric. For a realistic
scenario, one must include matter fields coupled to $g_{\mu
\nu}$, and possibly to $\phi$ also. However, with their
inclusion, the resulting equations of motion are difficult to
solve explicitly and the required analysis becomes very
involved. Hence, in the present paper, we will not include the
matter fields.

Writing the action given in (\ref{P action}) in terms of the
Einstein frame metric $g_{*\mu\nu}$ given by 
\be\label{weyl}
g_{*\mu\nu} = e^{\psi(\ph)}g_{\mu\nu},
\ee 
the action $S$ becomes
\be
S_* =  - {1\over{16\pi}} \int d^Dx {\sqrt{-g_*}} \left( 
R_{*} - {1\over2}(\nabla_{*}{\phi})^2 + V(\ph) \right)
+  S_{probe}(e^{-\psi}g_{*\mu\nu}) \; . \label{E action}
\ee
Written in terms of $g_{*\mu \nu}$, the action for the graviton
and scalar is in the canonical form. Hence, $g_{*\mu \nu}$ is
the Einstein frame metric and $\phi$ is the canonically
normalised scalar field. The probe however couples to $\phi$
also through the function $\psi(\ph)$. Indeed, this is our
definition of $\psi(\ph)$. This is why we get the rather unusual
coefficients for the scalar field terms in the action $S$ given
in (\ref{P action}). It is straightforward to write the action
$S$ equivalently in the generalised Brans-Dicke (BD) form
\cite{bd,rest,will}. Then, the corresponding BD function
$\omega_{BD}$ is given by
\be\label{wbd}
\omega_{BD} = {2\over{(D-2)^2{\psi}_{\ph}^2}} 
- {(D-1)\over(D-2)} \; . 
\ee
The equations of motion for $g_{* \mu \nu}$ and $\phi$ in the
Einstein frame are given by
\be
2R_{*\mu\nu} - {\nabla_{*\mu}}{\ph} {\nabla_{*\nu}}{\ph} 
+ {2\over{D-2}} V(\ph) & = & 0 \label{R eq in E} \\
{\nabla}_{*}^2{\ph} + {{\partial{V}}\over{\partial{\ph}}}
& = & 0 \; . \label{phi eq in E}
\ee 
The Ricci scalar $R$ in the physical frame is related to the
Ricci scalar $R_*$ in the Einstein frame by
\be
R = e^{\psi} \left( R_* + (D-1) (\psi_{\ph} \nabla_*^2 \ph 
+ \psi_{\ph\ph} (\nabla_*\ph)^2) 
- \frac{(D - 1) (D - 2)}{4} \psi_{\ph}^2 (\nabla_*\ph)^2 
\right) \; . \label{conR}
\ee
Other curvature invariants in the physical frame can also be
similarly related to those in the Einstein frame. Note that 
if $V =0$ then $R$ is given by 
\be
R & = & e^{\psi} R_* \left( 1 + 2 (D-1) \psi_{\ph\ph} 
- \frac{(D-1) (D-2)}{2} \psi^2_{\ph} \right) \; . \label{rr*}
\ee

We will solve the equations of motion in the Einstein
frame. Without matter included, solving the differential
equations of motion in this frame is much simpler. For getting
the physical quantities as seen by the probe we will transform
back to the physical frame. With matter included however,
solving the equations of motion is difficult in any frame. In
this paper, therefore, we consider only the vacuum solutions
with matter absent and study their physical properties as seen
by the probe. For some solutions with $V(\phi)$ and/or matter
present, see \cite{v,damour,kalyan}.

This procedure of obtaining solutions in one frame, here
Einstein frame, and then obtaining the physical quantities by
transformig back to the physical frame using $g_{\mu \nu} = e^{-
\psi} g_{* \mu \nu}$ (see equation (\ref{weyl})) is certainly
valid as long as the factor $e^{- \psi}$ is not zero or
infinity. If this factor vanishes (or diverges) at a point then
the validity of this procedure is not automatic; often, the
equivalence between the different frames will be destroyed at
this point. One then has to study the implications of the
transformed solutions in the physical frame, and check whether
they satisfy the equations of motion at this point also.
\footnote{We thank the referee for stressing the importance of
this point which we had overlooked earlier.} In fact, in the
case of a conformally coupled scalar field \cite{bek}, the
transformed solutions imply distributional sources at such a
point in the physical frame, as shown by a beautiful analysis in
\cite{sz}. We will comment more on this issue below.

\section{Cosmological Solutions}

Consider the FRW metric for flat universe in the physical
frame. It is given by
\be 
ds^{2} ~& \equiv &~ g_{\mu \nu} ~dx^{\mu}~dx^{\nu} \nonumber \\
& = &~ - dt^{2} ~+~ a^2(t)\left( dr^2  ~+~ r^{2}~
d\Omega^{2}_{D-2} \right)~, \label{frw} 
\ee
where $t$ and $a$ are the physical time and the scale factor as
seen by the probe. The corresponding metric in the Einstein
frame is given by
\be
ds_*^2 &=& g_{*\mu \nu} ~dx^{\mu}~dx^{\nu} \nonumber \\
       &=& -dt_*^2 + a_*^2(t_*)(dr^2 + r^2d\Omega^2_{D-2}) \; . 
\ee
Equation (\ref{weyl}) relates the cosmic times and the scale
factors in the two frames as follows: 
\be
{dt\over{dt_*}} = e^{-\psi /2}  &,&  \; \; \; 
a = a_* e^{-\psi /2} \; .  \label{at}
\ee
So, $t$ is a strictly increasing function of $t_*$ if $e^{-\psi
/2}>0$ strictly.

Now putting in the specific forms of the Ricci tensor components
the equations of motion (\ref{R eq in E}),(\ref{phi eq in E})
take the following form.
\be
\ddot{\ph} + (D-1){\dot{a_*}\dot{\ph}\over a_*} 
- {{\partial{V}}\over{\partial{\ph}}} & = & 0 \label{BDC1}\\
{\ddot{a_*}\over {a_*}} + (D-2) {{\dot{a_*}}^2\over {a_*^2}}
+ {1\over{D-2}} V(\ph) & = & 0 \label{BDC2}\\
2{(D-1)(D-2)}{{\dot{a_*}}^2\over {a_*^2}} 
- {\dot{\ph}}^2 + 2 V(\ph) & = & 0 \label{BDC3}
\ee
where $\dot{(\;)} \equiv \frac{d (\;)}{d t_*}$. The above
equations, with $V(\phi) \ne 0$, can be solved by the `method of
prepotentials' \cite{pre}. For a given $V(\phi)$ find the
prepotential $W(\phi)$ which satisfies the nonlinear
differential equation
\[
2 (D - 2)^2 W_\phi^2 - (D - 1) (D - 2) W^2 = V 
\]
where $W_\phi \equiv \frac{d W}{d \phi}$. Then $\phi(t_*)$ and
$a(t_*)$ can be obtained from the equations
\[
\dot{\phi} = - 2 (D - 2) W_\phi \; , \; \; \;
\frac{\dot{a}}{a} = W \; .
\]
The solutions thus obtained can be shown to satisfy the
equations (\ref{BDC1}) - (\ref{BDC3}). The difficulty in solving
these equations is now essentially transferred to solving for
the prepotential $W$ for a given $V$. Although solutions can be
obtained easily by this method for a class of potentials $V$, we
consider in this paper $V = 0$ case only. Our analysis can be
extended to $V \ne 0$ cases also.

So, let $V = 0$ in the following. Solving the equations of
motion, we get
\be
e^\phi & = & e^{\phi_0} \; \left( 
\frac{t_*}{t_{*0}} \right)^{\epsilon m} \label{cos phi} \\
a_* & = & a_{*0} \; 
\left( \frac{t_*}{t_{*0}} \right)^n \label{cos a}
\ee
where $\epsilon = \pm 1$, the range of $t_*$ is $0 \le t_* \le
\infty$, $m$ and $n$ are positive constants, and $a_* = a_{*0}$,
$\phi = \phi_0$ at some initial time $t_{*0} > 0$. The solution
to equations (\ref{BDC1}) - (\ref{BDC3}) gives
\be
(n,m) & = &  \left( {1\over{(D-1)}} \; , \;  
\sqrt{ \frac{2 (D-2)}{D-1} } \right) \; . 
\ee
The Ricci scalar $R_*$ then becomes
\be
R_* = - \left( {D-2\over{D-1}} \right) \; {1\over{t_*^2}} \; . 
\ee
The physical Ricci scalar $R$ is given by (\ref{rr*}). Note that
as $t_* \to 0$, $R_* \to \infty$.

To understand the behaviour of the physical quantities $t$, $a$,
$R$, and their dependence on the function $\psi(\phi)$, first
consider
\be
\psi(\ph) = k \ph
\ee
where $k$ is a constant which can be assumed to be positive
without loss of generality. Using equations (\ref{at}), and
defining $K \equiv 1 - \frac{\epsilon k m}{2}$, we get
\be
a & = & a_0 \; \left( \frac{t_*}{t_{*0}} 
\right)^{n - \epsilon k m/2}  \\
t - t_0 & = & \frac{B}{K} \; 
\left( \frac{t_*^K}{t_{*0}^K} - 1 \right) 
\; \; \; if \; \; \; K \ne 0 \nonumber \\ 
& = & B \; ln \left( \frac{t_*}{t_{*0}} \right)
\; \; \;   \; \; \; if \; \; \; K = 0 \\
R & = & C \; t_*^{- 2 K} 
\ee
where $a_0, B$, and $C$ are some constants whose explicit forms
are not needed and $t_0$ is the value of the physical time $t$
when $t_* = t_{*0}$. 

The evolution of $t, a$, and $R$ as $t_*$ varies from $0$ to
$\infty$ can be easily seen from the above equations. The values
of these quantities vary monotonically for $0 < t_* < \infty$.
Asymptotically, as $t_* \to 0$ or $\infty$, they tend to $0$,
$\pm \infty$, or to a finite value depending on whether $K$ is
positive, zero, or negative {\em i.e} whether $\epsilon k m$ is
less than, equal to, or greater than $2$. These asymptotic
behaviours, which can be obtained from the above equations, are
summarised conveniently in the Tables I and II given below.
\[
\begin{array}{||c||c|c|c||}
\hline \hline 
& t & a & |R| \\ \hline \hline
\epsilon k m < 2 &   finite & 0 \; or \; \infty  & \infty \\ \hline
\epsilon k m = 2 & - \infty & \infty             & finite \\ \hline
\epsilon k m > 2 & - \infty & \infty             & 0      \\ \hline
\hline
\end{array}
\]
\begin{center}
TABLE I: Asymptotic behaviour of $t, a$, and $|R|$ as $t_* \to
0$. \\ For $\epsilon k m < 2$, $a = 0$ or $\infty$ if $(2 n -
\epsilon k m)$ is positive or negative.
\end{center}
\[ 
\begin{array}{||c||c|c|c||} 
\hline \hline
 & t & a & |R| \\ \hline \hline
\epsilon k m < 2 & \infty & \infty \; or \; 0  & 0         \\ \hline
\epsilon k m = 2 & \infty & 0                  & finite     \\ \hline
\epsilon k m > 2 & finite & 0                  & \infty \\ \hline
\hline
\end{array}
\]
\begin{center}
TABLE II: Asymptotic behaviour of $t, a$, and $|R|$ as $t_* \to
\infty$. \\ For $\epsilon k m < 2$, $a = \infty$ or $0$ if $(2 n
- \epsilon k m)$ is positive or negative.
\end{center}

The behaviour of $t, a$, and $R$ as $t_*$ varies from $0$ to
$\infty$ can be read off from the above Tables. Generically
$\epsilon k m \ne 2$. If $\epsilon k m < 2$ then the time $t$
evolves to $\infty$ in future with no curvature singularity,
whereas in the past there is a curvature singularity at a finite
time $t$ beyond which the physical time $t$ cannot be extended.
The corresponding asymptotic values of the scale factor $a$
depends on the sign of $(2 n - \epsilon k m)$ and are given in
the Tables I and II. For $2 n - \epsilon k m > 0$, $a \to 0$ and
this is the usual Big Bang singularity encountered in general
relativity.  On the other hand, if $\epsilon k m > 2$ then the
past evolution is singularity free but the future evolution
terminates in a singularity at a finite time where the scale
factor vanishes - the Big Crunch. For $\epsilon k m = 2$, which
is non generic, the curvature scalar $R$ remains finite
throughout the past and the future.

We now consider the behaviour of the physical quantities $t, a,$
and $R$ for a general function $\psi(\phi)$. We will assume that
all the derivatives of $\psi$ with respect to $\phi$ are
finite. Namely,
\begin{equation}\label{k1}
\psi_{(n)}(\phi) \equiv \frac{d^n \psi}{d \phi^n} = finite 
\; \; \;  \forall n \ge 1. 
\end{equation}
This will ensure that the curvature scalar or other curvature
invariants will not diverge because of the divergences in
$\psi_{(n)}$ for some $n$. The qualitative features of the
evolution of $t, a$, and $R$ can be obtained easily using
equations (\ref{at}), (\ref{cos phi}), (\ref{cos a}), and the
Tables I and II. It is then clear that these quantities will
evolve smoothly as a function of $t_*$ for $0 < t_* < \infty$,
except perhaps in the asymptotic limit when $t_* = 0$ or
$\infty$ and $| \phi | \to \infty$. Therefore, we now analyse
the asymptotic behaviour of the solutions in this limit.
Consider the class of functions $\psi(\phi)$ where
\[
\psi(\phi) = - \lambda |\phi| 
\; \; {\rm as} \; \; |\phi| \to \infty
\]
and $\lambda$ is a positive constant. From the asymptotic
behaviour of the solutions given in the Tables I and II, it can
be seen by a straightforward analysis that if $\lambda m \ge 2$
then the physical time $t$ can be continued from $- \infty$ to
$\infty$, the scale factor $a$ will remain non zero, and the
curvature scalar will remain finite throughout the evolution.
Therefore, we assume that the function $\psi(\phi)$ satisfies
the constraint (\ref{k1}) and obeys the asymptotic condition
above with $\lambda m \ge 2$; {\em i.e.} that
\begin{equation}\label{k2}
\psi(\phi) = - \lambda |\phi| 
\; \; {\rm as} \; \; |\phi| \to \infty \; , \; \; \; 
\lambda \ge \sqrt{\frac{2 (D - 1)}{D - 2}} \; . 
\end{equation}

A wide class of functions $\psi(\phi)$ exists satisfying the
above properties. A simple example is $\psi(\ph) = - \lambda
\sqrt{ \phi^2 + c^2 }$ with $\lambda$ satisfying the condition
given in equation (\ref{k2}). Note also that, in the language of
generalised Brans-Dicke theories, this constraint on $\lambda$
and $\psi(\phi)$ translates into the following constraint on the
BD function $\omega_{BD}$:
\be\label{obd}
\omega_{BD} \le - {D \over {D-1} } < - 1 
\ee
which can be obtained from equation (\ref{wbd}). 

In obtaining the above results, it was implicitly assumed that
$e^{- \psi} > 0$ strictly. Otherwise, $t$ will not be a strictly
increasing function of $t_*$. It can now be seen that this
assumption is satisfied. This is because it follows that any
function $\psi(\phi)$ which satisfies the constraints given in
equations (\ref{k1}) and (\ref{k2}) will have a finite maximum,
namely $\psi \le \psi_{max} < \infty$. This then implies that
$e^{- \psi} > 0$ strictly.

The fact that the physical Ricci scalar remains finite and the
physical cosmic time can be extended into the past and future
without bound is indeed a remarkable result and indicates that
the Big bang singularity may be absent in the limited context
discussed here. The absence of singularity further requires all
curvature invariants in the physical frame to be finite. This
can indeed be shown to be true following the methods of
\cite{krama} for general $D-$dimensions for functions
$\psi(\phi)$ satisfying the constraints given above.

\section{The Schwarzschild solution}

We now consider static spherically symmetric solutions with
$V(\phi) = 0$. As before it is easier to solve the equations of
motion in the Einstein frame without matter incorporated. For
solutions with $V \ne 0$ and/or with matter included, see
\cite{v,damour}. Let the metric in the Einstein frame be given
by
\be
ds_*^2 & = & - f dt^2 + {dr^2 \over {g}} 
+ h^2 d{\Omega}^2_{D-2}
\ee
where $f, g$, and $h$ are functions of $r$. The equations of
motion are
\be
{2 f'' \over {f'}} - {f'\over {f}} + {g' \over {g}} 
+ 2 (D-2) {h' \over{h}} & = & 0 \label{SS1} \\
{h' g' \over{h g}} - {h 'f' \over {h f}} 
+ 2 {h'' \over{h}} + \frac{\phi'^2}{D - 2}  
& = & 0 \label{SS2'} \\
2 f g h h'' + f g' h h' + f' g h h' 
- 2 (D-3) f (1 - g h'^2) & = & 0 \label{SS3} \\
2 \phi'' + \left( {f' \over {f}} + {g' \over {g}} 
+ 2 (D-2) {h' \over {h}} \right) {\ph}' & = & 0 \label{SS4}
\ee
where $( \; )' \equiv \frac{d}{d r} (\; )$. Also, $R_* =
\frac{1}{2} g \phi'^2$ and the physical Ricci scalar $R$ is
given by equation (\ref{rr*}). Now the equations of motion can
be solved by taking the ansatz
\be
f   & = & Z^a\label{f}\\
g   & = & Z^b\label{g}\\
h^2 & = & r^2 Z^q\label{h}\\
e^\phi & = & e^{\phi_0} Z^p \label{phi sch} \\
where \; \; \; 
Z   & = &  1 - \left( {r_0\over{r}} \right)^{D-3} \label{Z}
\ee
and $\phi_0$ and $r_0$ are constants. With this ansatz, $R_* =
\frac{p^2}{2} Z'^2 Z^{b - 2}$ and the physical Ricci scalar $R$
is given by
\be 
R = \frac{p^2}{2} \; e^\psi Z'^2 Z^{b - 2} \; \left( 
1 + 2 (D - 1) \psi_{\ph\ph} - \frac{(D - 1) (D - 2)}{2} 
\psi^2_\phi \right) \; . 
\ee
Solving the equations of motion (\ref{SS1}) - (\ref{SS4}) gives
\be
a & = &  1 - (D - 3) q \label{a} \\
b & = & 1 - q  \label{b} \\
2 p^2 & = &  (D - 2) \; (2 - (D - 3) q) \; q \; . \label{constr}
\ee
Note that if $p = q = 0$ then $\phi = \phi_0$ and $a = b = 1$,
which is the standard Schwarzschild solution in $D-$dimensional
spacetime with its horizon at $r = r_0$.

From equation (\ref{constr}), it follows that $0 \le q \le
\frac{2}{D - 3}$ since $p^2 \ge 0$ and $D \ge 4$. For a given
$p$ there are two solutions for $q$, one on either side of
$\frac{1}{D - 3}$. We choose the branch where $0 \le q \le
\frac{1}{D - 3}$ so that when $p = 0$ one obtains $q = 0$ only,
{\em i.e.} the standard $D-$dimensional Schwarzschild solution
only with horizon located at $r_0$. The ranges of $q, a$ and
$b$ are thus
\[
0 \le q \le \frac{1}{D - 3} \; , \; \; \; 
1 \ge a \ge 0 \; , \; \; \; 
1 \ge b \ge \frac{D - 4}{D - 3}
\]
as can be obtained from equations (\ref{a}) and (\ref{b}). 

Consider the physical Ricci scalar $R$. It diverges at $r = r_0$
for $p \ne 0$. It also diverges at $r = 0$ about which we will
comment later. The divergence at $r_0$ is absent for the
standard Schwarzschild case where $p = 0$. The singular
behaviour of the curvature scalar at $r = r_0$, when $p \ne 0$,
is due to the piece $e^\psi Z^{b-2}$. So in order that there is
no singularity at $r_0$ we must have
\be
\lim_{|\phi| \to \infty} e^\psi Z^{b - 2} = finite
\ee
and $\psi_\phi$ , $\psi_{\phi \phi}$ must also be finite.

We wish to ensure that the Ricci scalar $R$ and other curvature
invariants do not diverge because of the divergences in the
derivatives of $\psi$ with respect to $\ph$. So here too we
impose (\ref{k1}) and consider the class of functions
$\psi(\phi)$ where $\psi(\phi) = - \lambda |\phi| \; \; {\rm as}
\; \; |\phi| \to \infty$. Consider now the behaviour of $R$ near
$r_0$ where $|\phi| \to \infty$. So, the above requirement
implies
\be
\lim_{|\phi| \to \infty} e^\psi Z^{b - 2} 
=  e^{-|\phi| \left( \la - {2 - b \over {|p|} } \right)} \; . 
\ee
This is finite if $\la \geq {2 - b \over {|p|}}$, that is if
\be
\la \geq {1 + q \over { \sqrt{ 
({D - 2 \over 2}) (2- (D - 3)q) q } } } \; . \label{sch in}
\ee
The right hand side of (\ref{sch in}) minimizes for $ q =
\frac{1}{D-2}$. Putting this minimum value in (\ref{sch in}),
we get
\be
\lambda & \geq & \sqrt{ {2 (D - 1) \over {D - 2}} } \; . 
\ee

This is the same constraint on $\la$ as obtained from the
cosmological case. Thus there is a solution with non zero $q$,
{\em i.e.} with a non trivial scalar field, and for which the
physical Ricci scalar $R$ does not diverge at $r = r_0$ whenever
$\psi(\phi)$ satisfies the constraint (\ref{k2}). The parameter
$q$ will lie in the range $0 < q_- \le q \le min \left(q_+,
\frac{1}{D - 3}\right)$ where $q_-$ and $q_+$ are the values of
$q$ saturating the inequality in (\ref{sch in}). Also, following
the methods of \cite{krama} for general $D-$dimensions, all
curvature invariants in the physical frame can be shown to be
finite at $r_0$ for functions $\psi(\phi)$ satisfying
constraints given in equations (\ref{k1}) and (\ref{k2}).

Let $\lambda = \sqrt{\frac{2 (D - 1)}{D - 2}}$. Then, the
solution with no divergence at $r_0$ is given by either $q = 0$
or $q = \frac{1}{D - 2}$. The former one is the standard
Schwarzschild solution. Consider the later one and the
corresponding metric. The solutions give
\[
a = q = \frac{1}{D - 2} \; , \; \; \; 
b = \frac{D - 3}{D - 2} \; , \; \; \; 
p^2 = \frac{\lambda^2}{4} \; . 
\] 
Substitute these values into the physical line element, assuming
that the function $\psi(\phi)$ satisfies the constraint
(\ref{k2}) with $\lambda = \sqrt{\frac{2 (D - 1)}{D - 2}}$.
Then $\psi(\phi) = - \lambda |\phi|$ near $r_0$ and we get
\begin{equation}\label{ads}
d s^2 = \frac{1}{Z} \; \left( - d t^2 
+ r^2 d \Omega_{D - 2}^2 \right) + \frac{d r^2}{Z^2} \; . 
\end{equation}
Let $r^{D - 3} \equiv r_0^{D - 3} (1 + \rho)$. Then $\rho \simeq
0$ near $r = r_0$ and the physical line element near $r_0$
becomes
\[
d s^2 \simeq \frac{1}{\rho} \; 
\left( - d t^2 + r_0^2 d \Omega_{D - 2}^2 \right) 
+ \frac{r_0^2}{(D - 3)^2} \; \frac{d \rho^2}{\rho^2} 
\]
which describes a $D-$dimensional anti de Sitter spacetime of
radius $\frac{r_0}{D - 3}$. This also shows that the curvature
invariants are not diverging near $r_0$. For a detailed analysis
of such metrics, see \cite{duff2}.

{\bf Radial Geodesic Motion of a Massive Probe}

Another interesting feature of the solution is the gravitational
force in the physical frame as seen by the probe. It is
attractive for $r \to \infty$ but becomes replusive for $r \to
r_{0+}$. To show this let us look at the $tt$ component of the
physical metric $g_{\mu\nu}$. Now, $g_{tt} = - e^{- \psi}
Z^a$. So, in the limit $r \to \infty$,
\be
- g_{tt} = 1 - (r_0/r)^{D-3} 
+ {\mathcal O}((r_0/r)^{2 (D-3)}) < 1 \; . 
\ee
$e^{-\psi}$ is non negetive and never vanishes since $\psi \le
\psi_{max} < \infty$. $Z>0$ for $r>r_0$ and
\be
\lim_{r \to r_0}(- g_{tt}) & = & \lim_{r \to r_0} 
Z^{a - \la |p|} \\
& \ge & \lim_{r \to r_0} Z^{a + b - 2} \rightarrow \infty \; . 
\ee
So as $r$ decreases from $\infty$ to $r_{0+}$, the above
mentioned factors ensure that ($-g_{tt}$) decreases from 1 to
some minimum value at $r_{min} > r_0$, and then diverges to
infinity as $r \to r_0$ always remaining positive and non
vanishing in this range. The slope of the curve of ($-g_{tt}$)
with respect to $r$ gives the nature of the gravitational
force. In the standard Schwarzschild solution, this is always
attractive. But the particular $r-$dependence here shows that
for scalar tensor theories the force is attractive for $r >
r_{min}$ and becomes replusive for $r < r_{min}$.

The repulsive force can be seen explicitly by studying the
geodesic motion of a radially incoming test particle with non
zero rest mass. For a metric given by
\bes
d s^2 = - g_0 d t^2 + g_1 d r^2 + g_2 d \Omega_{D-2}^2,
\ees
where $g_0, \; g_1$, and $g_2$ are functions of $r$ only, the
radial geodesic equation becomes
\be\label{acc}
r_{pp} + \frac{g'_1 r_p^2}{2 g_1} + \frac{g'_0}{2 g_1 g_0^2} = 0 \\ 
t_p = \frac{1}{g_0} 
\ee
where $(\;)' \equiv \frac{d (\;)}{d r}$ and 
$(\;)_p \equiv \frac{d (\;)}{d p}$. 
Equation (\ref{acc}) can be integrated twice to get 
\begin{equation}\label{tr}
\int dt = \int dr \sqrt{\frac{g_1}{g_0 (1 + E g_0)}} \; 
\end{equation}
where $E = - 1 + v^2$, corresponding to releasing the test
particle at $r = \infty$ with an inward velocity $v$ (in units
where velocity of light $= 1$). Since the test particle has non
zero rest mass, its velocity $v < 1$ and, hence, $E < 0$. 

In our case $g_0 = -{g}_{tt}$ which diverges to $\infty$ at
$r_0$.  Therefore, the denominator in (\ref{tr}) vanishes at
some $r_t$, where $1 + E g_0(r_t) = 0$ and $r_0 < r_t <
r_{min}$, indicating that $r_t$ is the turning point. Equation
(\ref{tr}) also shows that a test particle starting from $r =
r_{initial} < \infty$ reaches the turning point $r_t$ at a
finite physical time.  Analysis of equation (\ref{acc}) then
shows that the test particle travels outwards after reaching
$r_t$. It is clear that such a turning point exists irrespective
of the value of $v \; ( < 1)$ or, equivalently, the initial
energy of the test particle.  This shows that massive test
particles feel a repulsive gravitational force as they approach
$r_{0+}$. Contrast this with the Schwarzschild black hole where
$g_0 = 1 - { (\frac{r_0} {r}) }^{D-3} \le 1$: the factor $1 + E
g_0$ never vanishes and, hence, there is no turning point. In
Einstein frame, where the action is given by (\ref{E action})
and the test particle couples to the scalar field $\phi$ also,
this repulsion can be thought of as arising due to the the
scalar field force.

{\bf A Few Remarks}

We now make a few remarks about the properties of the static
spherically symmetric solutions obtained above. For $q = 0$ they
are the standard $D-$dimensional Schwarzschild solutions with
horizon located at $r_0$. Consider $q \ne 0$. \\
\noindent{\bf (i)}
In the Einstein frame, the curvature invariants diverge at
$r_0$. Hence, the above solutions do not describe black
holes. In the physical frame also, they do not describe black
holes because if $\psi(\phi)$ is unconstrained then the
curvature invariants diverge at $r_0$, whereas if they are
constrained then the physical frame component $g_{tt}$ does not
vanish for any $r$ where $r_0 \le r \le \infty$. Thus, clearly,
the parameter $q$ is not an extra `black hole hair' which is
forbidden by no hair theorems \cite{nohair}. \\
\noindent{\bf (ii)}
The above solutions are valid in the region $r > r_0$ where all
fields are regular. Solutions in the region $r < r_0$ can also
be obtained straightforwardly - either from the above solutions
with suitable modifications or otherwise. However, in all these
solutions, the scalar field $\phi$ will diverge as $r \to r_0$
in both the regions $r < r_0$ and $r > r_0$. Hence, these
solutions are likley to imply distributional sources at $r =
r_0$ as in \cite{sz}, in which case they can not be said to
satisfy the equations of motion (\ref{SS1}) - (\ref{SS4}) at $r
= r_0$.  \\
\noindent{\bf (iii)}
There is a singularity $r = 0$ in the region $r < r_0$ and it is
not removed even when $\psi$ satisfies the constraints
(\ref{k1}) and (\ref{k2}). \\
\noindent{\bf (iv)}
The physical metric $g_{\mu \nu} = e^{- \psi} g_{* \mu \nu}$.
For the functions $\psi$ satisfying the constraint (\ref{k2}),
the conformal factor $e^{- \psi}$ diverges at $r = r_0$.
Therefore, the equivalence of solutions in the Einstein and the
physical frame is likely to be destroyed at $r = r_0$. See also
the remark {\bf (ii)} above. \\
\noindent{\bf (v)}
In the physical frame, the curvature invariants are all finite
at $r = r_0$ when the function $\psi(\phi)$ satisfies the
constraints (\ref{k1}) and (\ref{k2}). It will be interesting to
extend, if possible, the present solutions across $r_0$. The
present coordinates are unlikely to be useful for such an
extension and one has to find a coordinate chart that can cover
appropriately the region around $r_0$. Note that when $\lambda =
\sqrt{\frac{2 (D - 1)}{D - 2}}$, one may effectively continue
across $r_0$ by lifting the solution to one higher dimension,
the details of which can be found in \cite{duff2}. But it is
not clear to us whether such an extension and interpretation as
in \cite{duff2} is possible for $\lambda > \sqrt{\frac{2 (D -
1)}{D - 2}}$ also.

These aspects concerning $r \le r_0$ are interesting and their
resolution is an important issue.  However, we will be concerned
here only with $r > r_0$. Also, conservatively, we take the
present solutions for $q \ne 0$ to be valid only in the region
$r_0 < r \le \infty$ since this suffices for our purposes
here. As we will argue in section V, for realistic cases with
matter fields included the solutions presented here and their
features will remain unchanged for $r > r_0$, whereas they will
be modified for $r \le r_0$ the modificatons being dependent on
the details of matter fields.

\section{Discussion and Conclusions}

We considered $D-$dimensional scalar tensor theories,
characterised by a function $\psi(\phi)$, and studied the
vacuum solutions where matter fields are absent. The test
particles are assumed to couple minimally, and only, to the
metric in the physical frame. Therefore, they follow the
geodesics of the physical frame metric and will probe the
properties of the corresponding spacetime backgrounds. Of
course, the motion of the probes will be invariant in any frame,
{\em e.g.} Einstein frame, but it will not be along the
geodesics of the corresponding metric since, generically, the
probe will couple to the scalar field also and feel a force due
to it \cite{fm,flanagan}.

We obtained vacuum cosmological and static spherically symmetric
solutions and found that they exhibit interesting features for a
class of theories where the function $\psi(\phi)$ satisfies the
constraints (\ref{k1}) and (\ref{k2}). For cosmological
solutions, the Ricci scalar remains finite and the time
continues indefinitely into the past and the future. Other
curvature invariants can also be shown to remain finite. So
these cosmological solutions are free of singularities.

For the static spherically symmetric case we obtained solutions
where the scalar field varies non trivially. If the scalar field
is constant then these solutions reduce to the standard
Schwarzschild ones, with horizon at $r = r_0$.  Otherwise, the
solutions have a new singularity at $r_0$ where the Ricci scalar
diverges. But the divergence at $r_0$ is absent when
$\psi(\phi)$ satisfies the constraints (\ref{k1}) and
(\ref{k2}). Other curvature invariants at $r_0$ can also be
shown to remain finite then. Also, a radially infalling probe
feels a repulsive gravitational force as it approaches $r_0$,
reaches a turning point $r_t > r_0$, and then travels
outwards. However, $r = 0$ remains singular, as in the
Schwarzschild solution, and will likely be seen by any
conformally coupled probe, {\em e.g.}  electromagnetic fields
(photons) in $D = 4$ spacetime. Also, a proper extension of the
solutions for $q \ne 0$ across $r_0$, satisfying the equations
of motion (\ref{SS1}) - (\ref{SS4}) for all $r$, is not clear to
us.

The above features are interesting. But, the most crucial issue
is whether such solutions can arise in realistic cases with
matter fields present. For conformally coupled matter, such as
electromagnetic fields in $D = 4$ spacetime with action of the
form $\int d^4 x \sqrt{- g} F_{\mu \nu} F^{\mu \nu}$, the
equations of motion for the metric, scalar, and matter fields
always admit the constant scalar solution. Quite possibly then
the asymptotic end points of the dynamics in these cases will be
indistinguishable from that in the standard Einstein theory with
no scalar present.

But, generically, the equations of motion in other cases will
not admit the constant scalar as a solution. Then, it is
possible that the soultions found here might describe the
asymptotic end points of the dynamics. For example, see
\cite{kalyan} for the $D = 4$ cosmological case with matter
included. A similar analysis, with similar conclusions, is
likely to be valid for $D > 4$ also.

Consider a `star' made up of, say, a perfect fluid which couples
minimally to the physical frame metric. Constant scalar is
indeed a solution outside the star in the vacuum but,
generically, not inside. Also, generically, the first derivative
of the scalar will be non zero at the boundary. Its continuity
across the boundary of the star would then imply that the vacuum
solution needs to be described by a solution with non trivial
scalar, namely a solution of the type presented here with $q \ne
0$ in the region $r > r_0$. The interior solution will depend on
the matter content and its distribution, and will be different
from that given here.

For the class of theories considered here, where $\psi(\phi)$
satisfies the constraints (\ref{k1}) and (\ref{k2}), the
gravitational force in the physical frame likely becomes
repulsive when the radius of a collapsing star approaches
$r_0$. Quite plausibly, this force will halt the collapse,
stabilising the star radius at a value near, but greater than,
$r_0$. If this is the case then the extension of the present
solutions, with $q \ne 0$, across $r_0$ will be rendered
unnecessary and the presence or absence of singularities in the
interior will be dictated by the matter content and its
distribution. The question of whether this is what actually
happens in a collapse in the class of scalar tensor theories
given here can only be answered by solving the relevent
equations of motion. But equations are very complicated and
solving them is beyond the scope of present paper. For some
solutions of `stars' in the scalar tensor theories, see
\cite{damour}. The scalar functions considered in these works,
however, do not satisfy the constraints (\ref{k2}).

Another important question is whether scalar tensor theories of
the type considered here can arise naturally. Scalar tensor
theories do appear naturally in various contexts {\em e.g.} in
Kaluza-Kelin (KK), string, and brane world theories
\cite{fm,rs,brane}.  Is the corresponding parameter $\lambda \ge
\sqrt{\frac{2 (D - 1)}{D - 2}}$; equivalently is $\omega_{BD}
\le - \frac{D}{D - 1} < - 1$?

It turns out that for KK theories $\omega_{BD} > - 1$
\cite{fm}. For low energy string theory $\omega_{BD} = - 1$
\cite{fm,polchinski}; however, see \cite{damourpolyakov}. In
string theory, if one considers the $D = 10$ spacetime probed by
the $D0-$brane probes then it turns out that the corresponding
$\omega_{BD} = - \frac{10}{9}$ \cite{duff} which is just on the
margin. In brane world theories of Randall-Sundrum I type
\cite{rs} having an extra fifth dimension of unit-interval
topology, the radion acts as the scalar in the effective four
dimensional theory on a brane located in the fifth dimension.
For the negative tension brane located at one end of the
interval, it turns out that $\omega_{BD} = - \frac{3}{2} +
\epsilon$ with $\epsilon$ positive and very close to zero
\cite{brane}. This value is $< - \frac{D}{D - 1}$ where $D = 4$
now. However, the implications are not completely clear to us
since in the brane world scenario, gravity propagates in the
fifth dimension and, moreover, the value of $\omega_{BD}$ on a
brane depends on its location in the fifth dimension. For
example, $\omega_{BD} = \infty$ on the positive tension brane,
located at the opposite end of the interval.

Another issue that needs to be studied is the following: if $D >
4$ and the observed four dimensional spacetime is to be part of
the $D-$dimensional spacetime in the scalar tensor theories
studied here then it is important to consider anisotropic cases
also since $(D - 4)$ directions are likely to be compact or, in
any case, have different dynamics from the observable four
dimensional spacetime.

Conservatively, we had taken here the static spherically
symmetric solutions for $q \ne 0$ to be valid only in the region
$r_0 < r \le \infty$. This was sufficient for our purposes
here. Nevertheless it is interesting and important in its own
right to study their extension across $r_0$, perhaps as in
\cite{duff2}, which satisfies the equations of motion for all
$r$, including $r_0$ and without any distributional sources as
in \cite{sz}.

\end{document}